\documentclass[conference]{article}

\usepackage{graphicx}
\usepackage[table,xcdraw]{xcolor}
\usepackage[hyphens]{url}
\usepackage{breakurl}
\usepackage[hidelinks]{hyperref}
\hypersetup{breaklinks=true}

\usepackage[color=yellow!60,textsize=footnotesize,obeyDraft,draft]{todonotes}
\usepackage{enumitem}
\usepackage{subcaption}
\usepackage{caption}
\usepackage{amssymb}
\usepackage{amsmath}
\usepackage{amsthm}
\usepackage{cleveref}

\usepackage{placeins}

\usepackage{multirow}
\usepackage{xurl} 

\usepackage{cite}
\usepackage{graphicx}

\begin{document}

\captionsetup[figure]{font={scriptsize,sf},labelformat={default},name={Fig.},labelsep=period}

\title{EnergyAnalyzer: Using Static WCET Analysis Techniques to Estimate the Energy Consumption of Embedded Applications}

\author{Simon Wegener\thanks{S. Wegener is with AbsInt, Saarbr\"ucken, Germany.
(e-mail: swegener@absint.com)}
, Kris K. Nikov\thanks{K. Nikov was with the University of Bristol, UK.
(e-mail: krisknikov@gmail.com)}
, Jose Nunez-Yanez\thanks{J. Nunez-Yanez is with Linköping University, Sweden.
(e-mail: jose.nunez-yanez@liu.se)}
\\and Kerstin Eder\thanks{K . Eder is with the University of Bristol, UK.
(e-mail: kerstin.eder@bristol.ac.uk)}
}%

\maketitle


\begin{abstract}
This paper presents \texttt{EnergyAnalyzer}, a code-level static analysis tool for estimating the energy consumption of embedded software based on statically predictable hardware events.
The tool utilises techniques usually used for worst-case execution time (WCET) analysis together with bespoke energy models developed for two predictable architectures---the ARM Cortex-M0 and the Gaisler LEON3---to perform energy usage analysis.
\texttt{EnergyAnalyzer} has been applied in various use cases, such as selecting candidates for an optimised convolutional neural network, analysing the energy consumption of a camera pill prototype, and analysing the energy consumption of satellite communications software.
The tool was developed as part of a larger project called TeamPlay, which aimed to provide a toolchain for developing embedded applications where energy properties are first-class citizens, allowing the developer to reflect directly on these properties at the source code level.
The analysis capabilities of \texttt{EnergyAnalyzer} are validated across a large number of benchmarks for the two target architectures and the results show that the statically estimated energy consumption has, with a few exceptions, less than 1\% difference compared to the underlying empirical energy models which have been validated on real hardware.
\end{abstract}


\section{Introduction}

Safety-critical embedded systems are used in various domains such as transportation, aerospace, medical devices, and industrial control systems. These systems are designed to meet certain non-functional requirements, such as timing or energy usage constraints, in addition to functional requirements. The satisfaction of these non-functional requirements is essential for the correct operation of the system and the safety of its users. Failure to meet these requirements can result in catastrophic consequences, such as loss of life or severe financial losses. Therefore, it is necessary to ensure that these systems are designed and implemented with reliable guarantees for their non-functional requirements.

For timing constraints, reliable guarantees can be obtained by using sound timing analysis methods.
Timing analysis is a technique used to analyse the temporal behaviour of a system and predict the worst-case execution time (WCET) of tasks and other timing properties.
Accurately determining a bound for the WCET of a task is essential for ensuring that a system meets its timing constraints and avoiding potential hazards.

Energy consumption is another crucial non-functional requirement for embedded systems.
Energy usage constraints are becoming more and more important due to the increasing use of battery-powered and energy-constrained devices.
Moreover, reducing energy consumption can increase the lifetime of the device and reduce its operating costs.
However, ensuring that a system meets its energy usage constraints is a challenging task, as energy consumption is highly dependent on the system's workload, input data, and hardware characteristics.
In contrast to timing analysis, which has a well-established theoretical foundation, creating an energy model that yields safe yet tight bounds for energy consumption is almost impossible. There are two primary reasons for this. First, energy consumption is measured in physical units (Joule), whereas processor cycles are a logical unit of time. Moreover, the amount of energy consumed by a processor is highly specific to the actual device. Two processors from the same production batch may already show a small difference in energy consumption. Additionally, the amount of energy consumed by one and the same processor may increase over time when the silicon inside the processor degrades~\cite{guo2018things}.
Second, the actual amount of energy consumed depends on the switching activity in the processor, which is highly data-dependent.
Thus, creating an energy model requires measuring all possible input combinations for each instruction, which is usually not feasible~\cite{Morse:2018}.
To address these limitations, several research works proposed using empirical methods to characterise energy models.
For example, Georgiou et al.~\cite{Georgiou:2017} suggest using pseudo-randomly created data to characterise an Instruction Set Architecture (ISA) energy model.
This approach reduces the number of input combinations needed to create the energy model and allows for faster evaluation of the model.

In recent years, researchers have proposed using event counters to create more accurate energy models for predictable architectures.
Event counters are hardware components that count the number of times certain events occur during execution, such as instructions executed or cache misses.
By using event counters to create an energy model, the model can accurately capture the energy consumption of a more diverse set of programs.
Furthermore, event counters are available on many modern processors, which makes the proposed energy models more accessible to developers.
Pallister et al.~\cite{pallister:2017} proposed an event counter-based method for data-dependent energy modelling, which is a more accurate way of modelling energy consumption for systems that process variable data sets. The proposed method identifies the relationship between the input data and the processor's energy consumption and uses this relationship to create an energy model. The authors evaluated their method on two different processors and found that it provided more accurate predictions of energy consumption than previous methods.

\texttt{EnergyAnalyzer}, presented in this paper, is a novel tool for static energy consumption analysis.
It focuses on estimating the energy consumption of programs running on two specific architectures: the Gaisler LEON3 microprocessor \cite{Leon3FT}, a radiation-tolerant microprocessor commonly used in the space communications sector, and the ARM Cortex-M0 microcontroller \cite{CortexM0}, known for its ultra-low power capabilities.
The key features of \texttt{EnergyAnalyzer} include:
\begin{enumerate}
    \item Utilising standard techniques from WCET analysis, for static worst-case energy consumption (WCEC) analysis.
    \item Incorporating accurate microarchitectural analysis and energy models~\cite{nikov2022accurate,nikov2022robust} for the two target architectures---ARM Cortex-M0 and Gaisler LEON3.
    \item When validated against model predictions using real-time samples, the static analysis shows <1\% difference in estimated energy for the vast majority of tested benchmarks.
\end{enumerate}


\section{Related Work}

Several studies have attempted to construct worst-case energy models capable of capturing the WCEC at the ISA level~\cite{Jayaseelan2006,wagemannworst}.
In Jayaseelan et al.~\cite{Jayaseelan2006}, the authors bound the WCEC on a simulated processor by maximizing the switching activity factor for each simulated component to obtain a WCEC cost for each ISA instruction.
Although this method retrieves the WCEC for all the ISA instructions, it could result in significant overestimation because the absolute worst-case on the hardware simulation used for the energy model's characterization phase might be infeasible to be triggered by any program on the actual hardware implementation of the same architecture.
Additionally, the approach is not feasible on physical hardware because there is no practical way of maximizing the switching activity on hardware.
To construct an equivalent ISA energy model for a fabricated processor, one would need to exhaustively search all combinations of valid data for the operands of an instruction, making it infeasible in most cases due to the huge space of possible input data combinations for each ISA instruction.
Although Wägemann et al.~\cite{wagemannworst} claimed to have constructed an energy model capable of capturing the WCEC, they did not provide the specifics of the model's characterization.
Nevertheless, they tested this energy model on a benchmark and admitted that such an absolute energy model could lead to significant overestimations, making the retrieved energy consumption bounds less useful.

Ideally, data-sensitive energy models would be created to capture the energy cost of executing an instruction based on the circuit switching activity caused by the operands used. Such models can potentially capture the WCEC of a program without overestimating the absolute energy consumption models. However, recent work has demonstrated that finding the data that triggers the WCEC is an NP-hard problem and that no practical method can approximate tight energy consumption upper bounds within any level of confidence~\cite{Morse:2018}. Therefore, the authors of Georgiou et al.~\cite{Georgiou:2017} suggested using pseudo-randomly created data to characterize an ISA energy model, as their empirical evidence showed that such models tend to be close to the actual worst case. Although this approach is expected to yield loose upper-bound energy consumption estimations, their experimental results showed a low level of underestimation of the WCEC (less than 4\%) for the programs tested. Such estimations can still provide valuable guidance to the application programmer to compare coding styles or algorithms in terms of resource consumption.

This work aims to estimate the WCEC using a similar approach to provide soft but tight energy bounds. Design decisions can be made based on empirical investigations to determine the level of over-provisioning that ensures the required level of dependability for the given application. This approach avoids the huge over-estimations and associated over-engineering at the system level that absolute worst-case energy models inherently exhibit.


\section{Tool Overview}

Over the last several years, a more or less standard architecture for static timing-analysis tools has emerged \cite{WEE+08, HWH95, Erm03, TFW00}, which is also implemented in AbsInt's WCET analyser \texttt{aiT}.
One can distinguish four major building blocks:
\begin{enumerate}
    \item The decoder translates the executable to an internal form that is used by the other parts (value analysis, microarchitectural analysis, etc).
    Architecture specific patterns decide whether an instruction is a call, branch, return, or just an ordinary instruction.
    This knowledge is used to form the basic blocks of the control-flow graph (CFG).
    Then, the control-flow between the basic blocks is reconstructed. In most cases, this is done completely automatically.
    However, if a target of a call or branch cannot be statically resolved, then the user needs to write some annotations to guide the control-flow reconstruction.

    \item Afterwards, the value analysis determines safe approximations of the values of processor registers and memory cells for every program point and execution context.
    These approximations are used to determine bounds on the iteration numbers of loops and information about the addresses of memory accesses.
    Value analysis information is also used to identify conditions that are always true or always false.
    Such knowledge is used to infer that certain program parts are never executed and therefore do not contribute to the worst-case resource consumption.
    Value analysis is again architecture-dependent.

    \item The microarchitectural analysis then determines upper bounds for the execution times of basic blocks by performing an abstract interpretation of the program execution on the particular architecture, taking into account its pipeline, caches, memory buses, and attached peripheral devices.
    The microarchitectural analysis is even more architecture-dependent than the decoder and value analysis, as the specification of the ISA alone does not suffice to create an abstract model of the hardware's timing behaviour, but the particular specifics of a particular processor implementing this specification must be taken into account (e.g., cache size, buffers, pre-fetching, etc).
    The microarchitectural analysis is usually a composition of both pipeline and cache analysis.

    \item Using the results of the preceding analysis phases, the path analysis phase searches for the worst-case execution path.
    The analysis translates the control-flow graph with the basic block timing bounds determined by the microarchitectural analysis and the loop bounds derived by the value analysis into an Integer Linear Program (ILP).
    The solution of the ILP yields a worst-case path together with a safe upper bound of the WCET.
    Path analysis is generic, i.e., does not depend on the target architecture.

\end{enumerate}
The structure of \texttt{EnergyAnalyzer} is similar to the structure of \texttt{aiT}.
In fact, both tools share most components.
In particular, they both use the same decoder for CFG reconstruction and the same value, loop, control-flow, and path analyses.
Only the microarchitectural analysis differs.
\texttt{aiT} uses a microarchitectural timing model to derive safe upper bounds of the WCET for each instruction.
In contrast, \texttt{EnergyAnalyzer} employs the microarchitectural energy models presented in Section~\ref{sec:energy_models} to estimate the energy consumption of each instruction.
%
Note, however, that for \texttt{EnergyAnalyzer}, the results of the path analysis have to be interpreted in a slightly different way, because the microarchitectural energy model of \texttt{EnergyAnalyzer} only produces an estimate of the WCEC.
Hence, during path analysis, worst-case execution frequencies of basic blocks are combined with an approximate energy model to produce a tight estimate of the worst-case energy behaviour which is not necessarily an upper bound.



\section{Microarchitectural Energy Analysis}
\label{sec:energy_models}

A key component of \texttt{EnergyAnalyzer} is the underlying use of accurate energy models for the target microarchitectures.
Several types of energy modelling techniques were explored and it was decided that hardware event-based power modelling is the most suitable approach, since it is a popular and accurate technique used for both CPU and full system modelling.
In their research, Rodrigues et al.~\cite{rodrigues2013study} conducted a systematic review of Performance Monitoring Unit (PMU) events also referred to as Performance Monitoring Counters (PMCs) commonly used in modern microprocessors.
They showed the effectiveness of these events in characterising and modelling dynamic power consumption.
Several other studies have also explored accurate power modelling~\cite{nunez2013enabling,rethinagiri2014system,Walker2017,seewald2021coarse,nikov2020intra}.

The PMC-based energy consumption estimation models were obtained via Ordinary Least Squares \cite{kutner2005applied} linear regression analysis, where coefficients, $\beta_x$, are determined for each counter, $C_x$, to predict the overall energy cost, i.e., $E = \sum_x (\beta_x \times C_x) + \alpha$, with $\alpha$ being the residual error term. The coefficients $\beta_x$ are the constants in the energy model that are program independent while the counters $C_x$ are the variables that depend on the program and its input. For a specific program with known counter values, the energy model can be used to estimate the energy consumed during the program’s execution.

For static-analysis-based energy consumption estimation, the overall energy consumption estimate of a piece of code is typically constructed from the estimates of the ISA basic blocks of the program. Thus, a PMC-based energy model can enable energy consumption estimation via static analysis only if the counters used for the modelling and prediction can be statically predicted at the ISA basic block level.
In order to make the model scalable for block-level static analysis, we have trained without using an intercept, so the residual is absorbed into the other event weights.
This means that at time 0 the energy predicted is also 0J.
We have also used a Non-Negative Least Squares (NNLS) solver to guarantee positive weights for all the events in the final energy model, thus always guaranteeing predictable positive energy consumption values from the model at discrete time slices~\cite{lawson1995solving}.

The accuracy of the model has been evaluated by using PMC data from a test set with the generated model equations.
The measured power or energy values are then compared to the estimations obtained from the model.
The percentage difference or mean absolute percentage error (MAPE) between them can be used as an objective metric to quantify model accuracy.
Several different models for each platform were identified and the best performing ones were integrated into \texttt{EnergyAnalyzer}.
Additional details on model generation and validation techniques used for both target platforms can be found in the accompanying papers~\cite{nikov2022robust,nikov2022accurate}.

\subsection{ARM Cortex-M0 Setup and Models}

The target platform on which the Cortex-M0 models were developed and validated is the STM32F0-Discovery board, which features the STM32F051 microprocessor~\cite{STMFO_technical}.
The platform does not feature an on-chip PMU.
Thus, a special methodology was developed to obtain the necessary PMC information, using an extended version of the \texttt{Thumbulator} instruction set simulator \cite{Thumbulator}.
The target platform allows ten different configurations, depending on CPU frequency, wait states for flash memory access, and whether instruction pre-fetch is enabled or not.
Further details on how the energy analysis models have been generated including a breakdown of the available hardware configurations and associated model weights and performance are presented in Nikov et al.~\cite{nikov2022accurate}.

We selected the following energy consumption model of the ARM Cortex-M0, using six statically predictable PMCs for integration into \texttt{EnergyAnalyzer}.
The model offers an estimation error to physical measurements, calculated as Mean Absolute Percentage Error (MAPE), of 2.8\%, for all the data points used for training and validation.
The resulting energy estimation is measured in nJ:

\begin{align*}
E_{\text{Cortex-M0}}
=&\; 0.972565030 \times C_{\text{executed instructions without multiplications}} \\
+&\; 0.652871770 \times C_{\text{RAM data reads}} \\
+&\; 1.031341343 \times C_{\text{RAM writes}} \\
+&\; 1.037625441 \times C_{\text{Flash data reads}} \\
+&\; 1.354953706 \times C_{\text{taken branches}} \\
+&\; 2.274650563 \times C_{\text{multiplication instructions}}
\end{align*}

\subsection{Gaisler LEON3 Setup and Models}

The LEON3 energy models were trained and validated on the GR712RC evaluation board~\cite{GR712RCDevBoard}. Similarly to the STM32F0-Discovery board, this platform also does not feature a PMU.
In order to get the PMC measurements for the models, a new, dual-platform approach using a Kintex UltraScale FPGA board was developed.
The programmable platform was loaded with a synthesised version of the LEON3 coupled together with the LEON3 Statistics Unit (L3STAT~\cite{grlib_ip}).
The results were synchronised with sensor measurements from the GR712RC platform to obtain the complete data set for model generation and validation.
More details on how the energy analysis models have been generated are presented in Nikov et al.~\cite{nikov2022robust}.

The models presented in that paper describe \textit{fine-grained} power models which are trained and validated on all available samples. The models integrated into \texttt{EnergyAnalyzer} use the same methodology, but with one key difference: the samples in the data set are aggregated for each benchmark to create code-block-sized models, making them more \textit{coarse-grained} and the NNLS solver is used to generate positive model weights, as required by \texttt{EnergyAnalyzer}.
Since average power models would not be very helpful for this purpose, total energy consumption is used instead. Additionally, the set of events that is supported by \texttt{EnergyAnalyzer} is limited, which has resulted in the creation of bespoke models for integration.
A list of all the supported PMCs can be found in Table~\ref{tab:energy:leon3:energyanalyser_pmu_events}.
A subset of them, labelled \textit{ISA+Cache} and shown in Table~\ref{tab:energy:leon3:isacache_pmu_events}, has been selected because it can be statically predicted with high accuracy.

Separate models are generated using each of the PMC subsets. In addition to using the \textit{bottom-up} and \textit{top-down} search algorithms, detailed in~\cite{nikov2022robust}, the relatively small PMC sets also allow for a \textit{full-exhaustive} search to be used. The resulting models are then compared against a model that uses all available PMCs. Table~\ref{tab:energy:leon3:coarse_grain_model_results} presents the results of the \textit{coarse-grained} model generation and evaluation.
As expected, the models computed using the larger \textit{All Supported} PMC list perform better than the ones using the \textit{ISA+Cache} list. However, it is interesting to note that the three search algorithms exploring the \textit{ISA+Cache} PMCs all converge on the same model with the \emph{STORE} event present in all models generated, regardless of PMC selection and search method. The  reason why the \textit{ISA+Cache} models perform worse than the \textit{All Supported} models is that the \emph{TIME} event, which is the single best predictor of power/energy according to previous work~\cite{nikov2015evaluation,nikov2020intra}, is not included in the list.
%
The \textit{ISA+Cache} energy model, with a MAPE of $<$8.3\%, was integrated into \texttt{EnergyAnalyzer} to provide energy estimations for the LEON3 in J.

\begin{table}
\centering
\resizebox{0.7\linewidth}{!}{%
\begin{tabular}{|r|l||r|l||r|l|}
\hline
\# & \textbf{Counter} & \# & \textbf{Counter} & \# & \textbf{Counter} \\ \hline
\hline
$C_{0}$  & TIME    & $C_{5}$  & WBHOLD  & $C_{14}$ & LDST    \\ \hline
$C_{1}$  & ICMISS  & $C_{7}$  & IINST   & $C_{15}$ & LOAD    \\ \hline
$C_{2}$  & ICHOLD  & $C_{11}$ & BRANCH  & $C_{16}$ & STORE   \\ \hline
$C_{3}$  & DCMISS  & $C_{12}$ & CALL    &    &         \\ \hline
$C_{4}$  & DCHOLD  & $C_{13}$ & TYPE2   &    &         \\ \hline
\end{tabular}%
}
\captionsetup{width=.8\textwidth}
\caption{\textit{All supported} PMCs by \texttt{EnergyAnalyzer}.}
\label{tab:energy:leon3:energyanalyser_pmu_events}
\end{table}

\begin{table}
\centering
\resizebox{0.7\linewidth}{!}{%
\begin{tabular}{|r|l||r|l||r|l|}
\hline
\# & \textbf{Counter} & \# & \textbf{Counter} & \# & \textbf{Counter} \\ \hline
\hline
$C_{1}$  & ICMISS  & $C_{11}$ & BRANCH  & $C_{14}$ & LDST    \\ \hline
$C_{3}$  & DCMISS  & $C_{12}$ & CALL    & $C_{15}$ & LOAD    \\ \hline
$C_{7}$  & IINST   & $C_{13}$ & TYPE2   & $C_{16}$ & STORE   \\ \hline
\end{tabular}%
}
\caption{\textit{ISA+Cache} subset of events.}
\label{tab:energy:leon3:isacache_pmu_events}
\end{table}

\begin{table}
\centering
\centering
\resizebox{0.9\linewidth}{!}{%
\begin{tabular}{|l||l||c|c|}
\hline
\multirow{2}{*}{Model}                                                                                                                                                                               & \multirow{2}{*}{Expression}                                                                                                                                                                                                                                                                                                                                                                                                                                                                                                                                                                                                                                                                                                                                                                               & \multicolumn{2}{c|}{MAPE{[}\%{]}} \\ \cline{3-4}
                                                                                                                                                                                                     &                                                                                                                                                                                                                                                                                                                                                                                                                                                                                                                                                                                                                                                                                                                                                                                                           & Train    & Test    \\ \hline \hline
\begin{tabular}[c]{@{}l@{}}Energy {[}J{]}\\ All Supported\\ All Events\end{tabular}      & \begin{tabular}[c]{@{}l@{}} $\text{E} = 0.155261 + 2.94155\text{e-}08\times{C_{0}}$\\$ + 	2.5661\text{e-}09\times{C_{2}} +	9.93453\text{e-}09\times{C_{5}}$\\$ +	8.97535\text{e-}10\times{C_{12}} +	3.21255\text{e-}09\times{C_{13}}$\\$ +	6.14384\text{e-}09\times{C_{15}} + 4.54827\text{e-}08\times{C_{16}}$\end{tabular}     &	1.14	& 0.29 \\ \hline
\begin{tabular}[c]{@{}l@{}}Energy {[}J{]}\\ All Supported\\ Bottom-Up\end{tabular}       & \begin{tabular}[c]{@{}l@{}} $\text{E} = 0	+	3.19557\text{e-}08\times{C_{0}}$\\$ +	5.79224\text{e-}08\times{C_{16}}$\end{tabular}     &	1.20	& 1.38 \\ \hline
\begin{tabular}[c]{@{}l@{}}Energy {[}J{]}\\ All Supported\\ Top-Down\end{tabular}        & \begin{tabular}[c]{@{}l@{}} $\text{E} = 0.131077	+	3.13122\text{e-}08\times{C_{0}}$\\$ +	9.17778\text{e-}09\times{C_{5}} +	2.99043\text{e-}09\times{C_{15}}$\\$ +	3.92999\text{e-}08\times{C_{16}} $\end{tabular}      &	1.02	& 1.54                          \\ \hline
\begin{tabular}[c]{@{}l@{}}Energy {[}J{]}\\ All Supported\\ Full-Exhaustive\end{tabular} & \begin{tabular}[c]{@{}l@{}} $\text{E} = 0.131087	+	3.13122\text{e-}08\times{C_{0}}$\\$ +	9.17779\text{e-}09\times{C_{5}} +	2.99043\text{e-}09\times{C_{14}}$\\$ +	3.63095\text{e-}08\times{C_{16}} $\end{tabular}    &	1.02	& 1.54                          \\ \hline
\begin{tabular}[c]{@{}l@{}}Energy {[}J{]}\\ IsaCache\\ All Events\end{tabular}           & \begin{tabular}[c]{@{}l@{}} $\text{E} = 0	+	1.18567\text{e-}06\times{C_{3}}$\\$ +	5.9072\text{e-}07\times{C_{12}} +	3.88949\text{e-}08\times{C_{13}}$\\$ + 8.03337\text{e-}08\times{C_{14}} +	6.89885\text{e-}08\times{C_{16}}$\end{tabular}      &	8.38	& 24.03 \\ \hline
\begin{tabular}[c]{@{}l@{}}Energy {[}J{]}\\ IsaCache\\ Bottom-Up\end{tabular}            & \begin{tabular}[c]{@{}l@{}} $\text{E} = 0	+	3.93365\text{e-}08\times{C_{7}}$\\$ +	1.87111\text{e-}07\times{C_{16}}$\end{tabular}     &	5.84	& 8.24 \\ \hline
\begin{tabular}[c]{@{}l@{}}Energy {[}J{]}\\ IsaCache\\ Top-Down\end{tabular}             & \begin{tabular}[c]{@{}l@{}} $\text{E} = 0	+	3.93365\text{e-}08\times{C_{7}}$\\$ +	1.87111\text{e-}07\times{C_{16}}$\end{tabular}     &	5.84	& 8.24 \\ \hline
\begin{tabular}[c]{@{}l@{}}Energy {[}J{]}\\ IsaCache\\ Full-Exhaustive\end{tabular}      & \begin{tabular}[c]{@{}l@{}} $\text{E} = 0	+	3.93365\text{e-}08\times{C_{7}}$\\$ +	1.87111\text{e-}07\times{C_{16}}$\end{tabular}     &	5.84	& 8.24 \\ \hline
\end{tabular}%
}
\caption{Coarse-grained Model Results}
\label{tab:energy:leon3:coarse_grain_model_results}
\end{table}

\newpage


\section{Evaluation}

We integrated the energy models from Section~\ref{sec:energy_models} into \texttt{EnergyAnalyzer for ARM Cortex-M0} and \texttt{EnergyAnalyzer for LEON3}, respectively.
We evaluated the integration with the help of the BEEBS benchmark suite~\cite{BEEBSb}.
The goal of the evaluation is to determine how close the statically estimated energy consumption for a given workload is to the model estimation.
Not all of the BEEBS benchmarks exercise the worst-case path through the program during execution.
Thus, a comparison of the results of the analysis and the actual measurements would compare in two orthogonal dimensions.
First, it would compare the tightness of the model with respect to the actual hardware measurements.
Second, it would compare the exercised path with the worst-case path.
In order to fix the comparison to one degree of freedom, we compared the energy estimates obtained from static analysis and those obtained from the energy model based on the actual PMC measurements from the platforms.
The tightness of the models has already been demonstrated in Section~\ref{sec:energy_models}.

In contrast to the safety-critical embedded hard real-time software that is usually analysed with \texttt{aiT}, the BEEBS benchmarks also contain dynamic memory management using \texttt{malloc} and \texttt{free}.
We did not analyse these benchmarks because the manual annotation effort to get tight results would be too high.
Some of the benchmarks contain computed calls via function pointers that cannot be resolved automatically.
In this case we manually annotated the call targets.
Moreover, we specified constant data in some cases.

For the LEON3, only a subset of the BEEBS benchmarks has been measured on the hardware setup, because the execution time of some of the benchmarks is too low to synchronise the FPGA and the ASIC (see Section~\ref{sec:energy_models} and accompanying paper~\cite{nikov2022robust}).

\subsection{EnergyAnalyzer for ARM Cortex-M0}
\label{subsec:energy_analyser_cm0}

For some benchmarks, the static analysis was not able to derive all loop bounds automatically.
In this case, we used \texttt{Thumbulator} to derive flow constraints for the ILP-based path analysis.
However, the benchmark might not exercise the worst-case path, and thus, using the simulation trace might not result in the worst-case amount of loop iterations for each loop in the program.
For one of the benchmarks---\emph{wikisort}---the simulation with \texttt{Thumbulator} fails because the binary allocated only 4096 bytes of stack, but the routine \texttt{WikiSort} already needs 4520 bytes of stack.
This causes a stack overflow.
Hence, some function pointer variables are overwritten, and the benchmark cannot be executed correctly.
We thus excluded the benchmark from the evaluation.
Two of the benchmarks---\emph{qsort} and \emph{select}---contain out-of-bounds accesses.

Table~\ref{tab:energy:eval:cortex-m0:beebs} shows the results of the evaluation of \texttt{EnergyAnalyzer for ARM Cortex-M0}.
For 43 benchmarks, the difference between the model and the static analysis is less than one percent, i.e., the execution path exercised during the simulator run is the worst-case path.
Note that the analysis results include the energy consumption of the execution of the main routine, which is not included in the simulator result, which only contains the path between the start trigger and the stop trigger.
However, the contribution of this overhead is less than one mJ and hence, negligible.
For the other benchmarks, the static analysis selected different paths as worst-case execution paths.
The maximal observed difference between the simulator run and the static analysis is 109\% for benchmark \emph{nsichneu}, which models a state machine with many different execution paths, and the static analysis was not able to prune infeasible paths.
Since the path analysis is a worst-case analysis, it maximises over the possible execution paths.
Hence, the path analysis selects the worst-case combination which differs significantly from the simulated execution path.

\texttt{EnergyAnalyzer} allows to trade performance for precision by specifying how many calling and loop contexts should be distinguished during analysis.
We used this feature to increase the analysis precision.
The analysis of most benchmarks takes less than four minutes to complete, with the exception of five benchmarks (\emph{rijndael}, \emph{cubic}, \emph{sqrt}, \emph{nbody}, \emph{picojpeg}), which took between 4 and 59 minutes.

\begin{table}[hp!]
\begin{minipage}{\linewidth}
\centering
\resizebox{\textwidth}{!}{%
\begin{tabular}{|c|c|c|r|c|}
	\hline
\textbf{Benchmark}   & \textbf{Analysis Result}              & \textbf{Model Result}                 & \textbf{$\Delta$} & \textbf{Note}\\ \hline\hline
aha-compress         & \phantom{0}78.885\makebox[2em][c]{mJ} & \phantom{0}78.828\makebox[2em][c]{mJ} &          <\,1\,\% & \\ \hline
aha-mont64           & \phantom{0}99.396\makebox[2em][c]{mJ} & \phantom{0}99.396\makebox[2em][c]{mJ} &          <\,1\,\% & \\ \hline
bubblesort           &           366.763\makebox[2em][c]{mJ} &           366.762\makebox[2em][c]{mJ} &          <\,1\,\% & \\ \hline
cnt                  & \phantom{0}42.813\makebox[2em][c]{mJ} & \phantom{0}42.804\makebox[2em][c]{mJ} &          <\,1\,\% & \\ \hline
compress             & \phantom{0}27.895\makebox[2em][c]{mJ} & \phantom{0}27.895\makebox[2em][c]{mJ} &          <\,1\,\% & \\ \hline
crc                  & \phantom{00}9.623\makebox[2em][c]{mJ} & \phantom{00}9.623\makebox[2em][c]{mJ} &          <\,1\,\% & \\ \hline
cubic                & \phantom{00}7.801\makebox[2em][c]{J}  & \phantom{00}4.138\makebox[2em][c]{J}  &            89\,\% & flow constraints \\ \hline
duff                 & \phantom{00}4.349\makebox[2em][c]{mJ} & \phantom{00}4.349\makebox[2em][c]{mJ} &          <\,1\,\% & \\ \hline
edn                  &           302.762\makebox[2em][c]{mJ} &           302.762\makebox[2em][c]{mJ} &          <\,1\,\% & \\ \hline
expint               & \phantom{0}43.315\makebox[2em][c]{mJ} & \phantom{0}43.315\makebox[2em][c]{mJ} &          <\,1\,\% & \\ \hline
fac                  & \phantom{00}2.934\makebox[2em][c]{mJ} & \phantom{00}2.904\makebox[2em][c]{mJ} &             1\,\% & \\ \hline
fasta                & \phantom{0}29.383\makebox[2em][c]{J}  & \phantom{0}21.100\makebox[2em][c]{J}  &            39\,\% & flow constraints \\ \hline
fdct                 & \phantom{0}12.292\makebox[2em][c]{mJ} & \phantom{0}12.292\makebox[2em][c]{mJ} &          <\,1\,\% & \\ \hline
fibcall              & \phantom{00}1.493\makebox[2em][c]{mJ} & \phantom{00}1.493\makebox[2em][c]{mJ} &          <\,1\,\% & \\ \hline
fir                  & \phantom{00}1.994\makebox[2em][c]{J}  & \phantom{00}1.994\makebox[2em][c]{J}  &          <\,1\,\% & \\ \hline
frac                 & \phantom{00}1.183\makebox[2em][c]{J}  & \phantom{00}1.183\makebox[2em][c]{J}  &          <\,1\,\% & \\ \hline
insertsort           & \phantom{00}3.089\makebox[2em][c]{mJ} & \phantom{00}3.089\makebox[2em][c]{mJ} &          <\,1\,\% & \\ \hline
janne\_complex       & \phantom{00}1.402\makebox[2em][c]{mJ} & \phantom{00}1.402\makebox[2em][c]{mJ} &          <\,1\,\% & \\ \hline
jfdctint             & \phantom{0}31.481\makebox[2em][c]{mJ} & \phantom{0}31.476\makebox[2em][c]{mJ} &          <\,1\,\% & \\ \hline
lcdnum               &           886.941\makebox[2em][c]{uJ} &           805.000\makebox[2em][c]{mJ} &            10\,\% & \\ \hline
levenshtein          &           400.926\makebox[2em][c]{mJ} &           400.926\makebox[2em][c]{mJ} &          <\,1\,\% & \\ \hline
ludcmp               &           174.559\makebox[2em][c]{mJ} &           174.559\makebox[2em][c]{mJ} &          <\,1\,\% & \\ \hline
matmult-float        & \phantom{00}1.537\makebox[2em][c]{J}  & \phantom{00}1.537\makebox[2em][c]{J}  &          <\,1\,\% & \\ \hline
matmult-int          &           842.724\makebox[2em][c]{mJ} &           842.649\makebox[2em][c]{mJ} &          <\,1\,\% & \\ \hline
minver               &           131.316\makebox[2em][c]{mJ} & \phantom{0}84.348\makebox[2em][c]{mJ} &            56\,\% & flow constraints \\ \hline
nbody                & \phantom{0}25.844\makebox[2em][c]{J}  & \phantom{0}25.844\makebox[2em][c]{J}  &          <\,1\,\% & \\ \hline
ndes                 &           293.387\makebox[2em][c]{mJ} &           293.297\makebox[2em][c]{mJ} &          <\,1\,\% & \\ \hline
nettle-arcfour       &           105.880\makebox[2em][c]{mJ} &           105.880\makebox[2em][c]{mJ} &          <\,1\,\% & \\ \hline
nettle-cast128       & \phantom{0}23.214\makebox[2em][c]{mJ} & \phantom{0}23.211\makebox[2em][c]{mJ} &          <\,1\,\% & \\ \hline
nettle-des           & \phantom{0}22.595\makebox[2em][c]{mJ} & \phantom{0}22.595\makebox[2em][c]{mJ} &          <\,1\,\% & \\ \hline
nettle-md5           & \phantom{00}5.467\makebox[2em][c]{mJ} & \phantom{00}5.467\makebox[2em][c]{mJ} &          <\,1\,\% & \\ \hline
nettle-sha256        & \phantom{0}50.507\makebox[2em][c]{mJ} & \phantom{0}50.507\makebox[2em][c]{mJ} &          <\,1\,\% & \\ \hline
newlib-exp           & \phantom{0}70.439\makebox[2em][c]{mJ} & \phantom{0}70.439\makebox[2em][c]{mJ} &          <\,1\,\% & \\ \hline
newlib-log           & \phantom{0}52.954\makebox[2em][c]{mJ} & \phantom{0}52.954\makebox[2em][c]{mJ} &          <\,1\,\% & \\ \hline
newlib-sqrt          & \phantom{0}10.289\makebox[2em][c]{mJ} & \phantom{0}10.289\makebox[2em][c]{mJ} &          <\,1\,\% & \\ \hline
nsichneu             & \phantom{0}61.017\makebox[2em][c]{mJ} & \phantom{0}29.185\makebox[2em][c]{mJ} &           109\,\% & \\ \hline
picojpeg             & \phantom{00}4.885\makebox[2em][c]{J}  & \phantom{00}4.885\makebox[2em][c]{J}  &          <\,1\,\% & \\ \hline
prime                &           209.663\makebox[2em][c]{mJ} &           209.663\makebox[2em][c]{mJ} &          <\,1\,\% & \\ \hline
qsort                & \phantom{0}27.294\makebox[2em][c]{mJ} & \phantom{0}20.408\makebox[2em][c]{mJ} &            34\,\% & flow constraints \\ \hline
qurt                 &           139.891\makebox[2em][c]{mJ} &           139.890\makebox[2em][c]{mJ} &          <\,1\,\% & \\ \hline
rijndael             & \phantom{00}7.176\makebox[2em][c]{J}  & \phantom{00}7.042\makebox[2em][c]{J}  &             2\,\% & \\ \hline
sglib-arraybinsearch & \phantom{0}76.596\makebox[2em][c]{mJ} & \phantom{0}76.596\makebox[2em][c]{mJ} &          <\,1\,\% & \\ \hline
sglib-arrayheapsort  & \phantom{0}86.857\makebox[2em][c]{mJ} & \phantom{0}86.857\makebox[2em][c]{mJ} &          <\,1\,\% & \\ \hline
sglib-arrayquicksort & \phantom{0}65.600\makebox[2em][c]{mJ} & \phantom{0}65.600\makebox[2em][c]{mJ} &          <\,1\,\% & \\ \hline
sglib-queue          &           126.250\makebox[2em][c]{mJ} &           126.250\makebox[2em][c]{mJ} &          <\,1\,\% & \\ \hline
slre                 &           206.734\makebox[2em][c]{mJ} &           206.734\makebox[2em][c]{mJ} &          <\,1\,\% & \\ \hline
sqrt                 & \phantom{0}11.529\makebox[2em][c]{J}  & \phantom{0}11.529\makebox[2em][c]{J}  &          <\,1\,\% & \\ \hline
st                   & \phantom{00}4.142\makebox[2em][c]{J}  & \phantom{00}2.945\makebox[2em][c]{J}  &            41\,\% & flow constraints \\ \hline
statemate            & \phantom{0}13.331\makebox[2em][c]{mJ} & \phantom{00}9.308\makebox[2em][c]{mJ} &            43\,\% & \\ \hline
stb\_perlin          & \phantom{00}5.145\makebox[2em][c]{J}  & \phantom{00}5.145\makebox[2em][c]{J}  &          <\,1\,\% & \\ \hline
stringsearch1        & \phantom{0}46.362\makebox[2em][c]{mJ} & \phantom{0}46.362\makebox[2em][c]{mJ} &          <\,1\,\% & \\ \hline
strstr               & \phantom{00}5.480\makebox[2em][c]{mJ} & \phantom{00}5.480\makebox[2em][c]{mJ} &          <\,1\,\% & \\ \hline
trio-snprintf        &           105.378\makebox[2em][c]{mJ} & \phantom{0}65.427\makebox[2em][c]{mJ} &            61\,\% & flow constraints \\ \hline
trio-sscanf          &           139.345\makebox[2em][c]{mJ} & \phantom{0}71.618\makebox[2em][c]{mJ} &            95\,\% & flow constraints \\ \hline
ud                   & \phantom{0}21.863\makebox[2em][c]{mJ} & \phantom{0}21.862\makebox[2em][c]{mJ} &          <\,1\,\% & \\ \hline
whetstone            & \phantom{0}22.533\makebox[2em][c]{J}  & \phantom{0}16.687\makebox[2em][c]{J}  &            35\,\% & flow constraints \\ \hline
\end{tabular}
}
\label{tab:energy:eval:cortex-m0:beebs2}
\end{minipage}
\caption{Evaluation of the integration of the energy model for the ARM Cortex-M0 into static energy consumption analysis.}
\captionsetup{width=.8\textwidth}
\label{tab:energy:eval:cortex-m0:beebs}
\end{table}

\newpage

\subsection{EnergyAnalyzer for LEON3}
\label{subsec:energy_analyser_leon3}

Some of the BEEBS benchmarks contain floating-point computations.
However, since the FPGA implementation of the LEON3 was built without a FPU, the benchmarks cannot use floating-point instructions but must use a software library that emulates these floating-point computations.
One of the benchmarks---\emph{minver}---computes a matrix multiplication using floating-point numbers, where one of the matrices is never initialised.
We performed both the standard worst-case analysis for this benchmark and an analysis where we assumed that the computations only process normalised IEEE~754 floating-point numbers and zero.
This reflects the ``flush to zero'' option present in many architectures.
The computed energy consumption estimate is then cut in half, which shows that enabling ``flush to zero'' in software floating-point computations can save a lot of energy.

The LEON3 implements the SPARCv8 ISA, which uses register windows for fast context switches, and for providing hardware support for the call stack.
However, the number of register windows is limited.
The particular LEON3 model used for our experiments, available on the the GR712RC board and its FPGA equivalent, has eight register windows.
Due to the overlapping nature of the register windows, and their use as a ring buffer, only seven are usable.
Hence, in case a program needs more than seven register windows, the processor triggers software traps to handle the register window overflow (and underflow).
This happens for two of the benchmarks---\emph{picojpeg} and \emph{slre}.
Hence, the processor needs to execute trap functions when it detects a register window overflow or a register window underflow.
This causes additional energy consumption which must be taken into account during a system-level energy analysis.

Table~\ref{tab:energy:eval:leon3:beebs} shows the results of the evaluation of \texttt{EnergyAnalyzer for LEON3}.
The analysis of most benchmarks takes less than four minutes to complete, but for three benchmarks---\emph{matmult-float}, \emph{nbody}, and \emph{picojpeg}---the analysis duration was 50 minutes, 45 minutes, and 24 minutes, respectively.

\begin{table}[hp!]
\begin{minipage}{1\linewidth}
\centering
\resizebox{\textwidth}{!}{%
\begin{tabular}{|c|c|c|r|c|}
\hline
\textbf{Benchmark}   & \textbf{Analysis Result} & \textbf{Model Result} & \textbf{$\Delta$} & \textbf{Note}\\ \hline\hline
aha-compress         & \phantom{0}11.004 J  & \phantom{0}11.004 J  &             0\,\% & \\ \hline
aha-mont64           & \phantom{00}7.499 J  & \phantom{00}7.491 J  &          <\,1\,\% & \\ \hline
bubblesort           & \phantom{00}3.898 J  & \phantom{00}3.889 J  &          <\,1\,\% & \\ \hline
edn                  & \phantom{0}39.186 J  & \phantom{0}39.186 J  &             0\,\% & \\ \hline
fir                  &           159.469 J  &           159.469 J  &             0\,\% & \\ \hline
frac                 & \phantom{0}59.391 J  & \phantom{0}59.339 J  &          <\,1\,\% & \\ \hline
levenshtein          & \phantom{0}25.506 J  & \phantom{0}25.491 J  &          <\,1\,\% & \\ \hline
ludcmp               & \phantom{0}10.992 J  & \phantom{0}10.814 J  &             2\,\% & \\ \hline
matmult-float        & \phantom{00}2.847 J  & \phantom{00}2.822 J  &             1\,\% & \\ \hline
minver               & \phantom{0}14.372 J  & \phantom{00}4.643 J  &           210\,\% & worst-case \\ \hline
minver               & \phantom{00}7.398 J  & \phantom{00}4.643 J  &            59\,\% & assumptions \\ \hline
nbody                & \phantom{00}4.512 J  & \phantom{00}4.496 J  &          <\,1\,\% & \\ \hline
ndes                 & \phantom{0}24.828 J  & \phantom{0}24.467 J  &             1\,\% & \\ \hline
nettle-aes           & \phantom{0}19.401 J  & \phantom{0}19.389 J  &          <\,1\,\% & \\ \hline
nettle-arcfour       & \phantom{00}9.644 J  & \phantom{00}9.639 J  &          <\,1\,\% & \\ \hline
nettle-sha256        & \phantom{00}2.763 J  & \phantom{00}2.754 J  &          <\,1\,\% & \\ \hline
newlib-exp           & \phantom{00}4.374 J  & \phantom{00}4.319 J  &             1\,\% & \\ \hline
newlib-log           & \phantom{00}3.284 J  & \phantom{00}3.252 J  &             1\,\% & \\ \hline
picojpeg             &           503.732 J  &           503.918 J  &         <\,-1\,\% & traps \\ \hline
prime                & \phantom{00}3.670 J  & \phantom{00}3.667 J  &          <\,1\,\% & \\ \hline
qurt                 & \phantom{00}8.001 J  & \phantom{00}7.958 J  &             1\,\% & \\ \hline
sglib-arraybinsearch & \phantom{00}6.283 J  & \phantom{00}6.281 J  &          <\,1\,\% & \\ \hline
sglib-arrayheapsort  & \phantom{0}13.066 J  & \phantom{0}13.062 J  &          <\,1\,\% & \\ \hline
sglib-arrayquicksort & \phantom{0}13.066 J  & \phantom{0}13.052 J  &          <\,1\,\% & \\ \hline
sglib-queue          & \phantom{0}13.901 J  & \phantom{0}13.900 J  &          <\,1\,\% & \\ \hline
slre                 & \phantom{0}14.988 J  & \phantom{0}15.261 J  &            -2\,\% & traps \\ \hline
\end{tabular}
}
\end{minipage}
\caption{Evaluation of the integration of the \emph{ISA+Cache} energy model for the LEON3 into static energy consumption analysis.}
\label{tab:energy:eval:leon3:beebs}
\end{table}

\newpage


\section{Integration into TeamPlay Toolchain and Case Studies}

\texttt{EnergyAnalyzer for ARM Cortex-M0} and \texttt{EnergyAnalyzer for LEON3} can be used as standalone tools to estimate the energy consumption of embedded software.
They provide a rich and user-friendly graphical user interface to ease the analysis process.
However, they have been developed during the TeamPlay project as part of a larger toolchain where they enable multi-criteria optimisation in a compiler, contract-based programming, and energy-aware scheduling.
In the following, we present the integration of \texttt{EnergyAnalyzer} into the WCET-aware C compiler \texttt{WCC}~\cite{falk2010compiler}.

The mechanisms implemented to integrate \texttt{EnergyAnalyzer} within \texttt{WCC} mirror the mechanisms in place to perform WCET analysis using AbsInt's WCET analyser \texttt{aiT}.
XTC files \cite{AbsIntXTC} are used to call \texttt{aiT} and \texttt{EnergyAnalyzer} in batch mode (i.e., without graphical user interface).
An XTC file specifies the binary to be analyzed, the entry point, the path to an annotation file which contains details about the target architecture configuration as well as user-provided annotations like flow facts, and the path to an XML report file.
This XML output file is then parsed by \texttt{WCC} after invocation of \texttt{EnergyAnalyzer} to extract the analysis results and import them into \texttt{WCC}'s Low-Level IR at function and basic block level.
This attached energy data can further be exploited by \texttt{WCC} to perform various compiler-level energy-aware optimisations, and thus, establishing a smooth flow between compiler-level energy analysis and optimisation.
\texttt{WCC} supports source-level flow facts utilizing ANSI C pragmas.
\texttt{WCC} uses these pragmas to direct integrated analysers during analysis.
In general, a user can annotate their code with loop bounds, recursion depths, and execution frequency of an instruction relative to some other instruction.
These source-level pragmas are translated within \texttt{WCC} into AIS2 annotations for \texttt{aiT} and \texttt{EnergyAnalyzer}.

\texttt{EnergyAnalyzer} has been applied to several use cases in the course of the TeamPlay project.
First, it has been used to select candidates from a set of implementations of compute elements for an optimised convolutional neural network (CNN).
Acting as an evaluation guide, it helped decide which optimisations should be considered for the final CNN implementation and which showed unacceptable energy consumption and should not be used.
Second, it has been used to analyse the energy consumption of a camera pill prototype.
The addition of an encryption algorithm showed a significant impact on the energy consumption of the camera pill that also varied depending on the type of encryption algorithm, with SPECK being an order of magnitude more energy efficient than AES and PRESENT.
\texttt{EnergyAnalyzer} closely predicted the actual energy usage that was physically measured on the system, and thus is a viable method for estimating energy consumption of a system such as the camera pill.
Third, \texttt{EnergyAnalyzer} has been used to analyse the energy consumption of a piece of satellite software.
One of the main challenges of the space industry is power consumption, as spacecrafts usually have limited access to power sources.
\texttt{EnergyAnalyzer} proved to be a useful tool thanks to its support of the LEON3 on the GR712RC platform, which is the most common processor ASIC used by European space companies.
The results showed a precise prediction of the execution energy for the different binaries.
Another interesting feature of \texttt{EnergyAnalyzer} was the result visualisation using call and control-flow graphs, showing the energy consumption for each of the functions inside the binary which could be used not only for predicting and minimising energy consumption but also for qualifying code for space.

More details on the integration of \texttt{EnergyAnalyzer} in the TeamPlay toolchain and the use cases on which the toolchain has been applied are presented in Rouxel et al.~\cite{rouxel2022teamplay}.


\section{Conclusion}

This paper presents our work on static energy consumption analysis for embedded systems.
We created energy models for two predictable architectures: the ARM Cortex-M0 and the Gaisler LEON3, both achieving estimation errors of less than 10\% when validated against our two target hardware platforms.
Both models are based on hardware event counters, which can be predicted by static analysis.
The models are integrated into \texttt{EnergyAnalyzer}, a novel tool for code-level static energy consumption analysis.
Our evaluation results show a good accuracy of the energy consumption analysis.
The tool provides a user-friendly graphical interface to analyse energy consumption at different levels of granularity, from the entire program to individual functions and basic blocks.
\texttt{EnergyAnalyzer} is part of the TeamPlay toolchain~\cite{rouxel2022teamplay}, where it enables multi-criteria code optimisation during compilation.

We demonstrated the usefulness of our tools in several case studies including a camera-pill designed for medical diagnosis and a space communications platform. In each case, the tool provided precise predictions of the energy consumption and helped to identify energy bottlenecks.
In conclusion, our work provides a useful approach to analyse energy consumption in embedded systems.
A potential avenue for extension is to include peripheral energy consumption for system-level analysis~\cite{wagemann2018whole}.
We believe that our approach can help to design more energy-efficient embedded systems and applications, which is a crucial step towards sustainable development.

\section*{Acknowledgement}

This research was supported by the European Union's Horizon 2020 Research and Innovation Programme under grant agreement No. 779882, TeamPlay (Time, Energy and security Analysis for Multi/Many-core heterogeneous PLAtforms).


\bibliography{shortedRef}

\newcommand{\etalchar}[1]{$^{#1}$}
\begin{thebibliography}{GVGGS18}

\bibitem[Abs]{AbsIntXTC}
{XTC Language Specification, Version 2.7}.
\newblock \url{https://www.absint.com/xtc/xtc-specification.pdf}.
\newblock Accessed: 2020-01-13.

\bibitem[{Cob}19]{grlib_ip}
{Cobham Gaisler AB}.
\newblock {GRLIB IP Core User's Manual}, 2019.
\newblock \url{https://www.gaisler.com/products/grlib/grip.pdf}.

\bibitem[Cor]{CortexM0}
{The ARM Cortex-M0 processor}.
\newblock
  \url{https://developer.arm.com/products/processors/cortex-m/cortex-m0}.
\newblock Accessed: 2018-08-14.

\bibitem[Erm03]{Erm03}
Andreas Ermedahl.
\newblock {\em A Modular Tool Architecture for Worst-Case Execution Time
  Analysis}.
\newblock PhD thesis, Uppsala University, Sweden, 2003.

\bibitem[FL10]{falk2010compiler}
Heiko Falk and Paul Lokuciejewski.
\newblock A compiler framework for the reduction of worst-case execution times.
\newblock {\em Real-Time Systems}, 46(2):251--300, 2010.

\bibitem[GKCE17]{Georgiou:2017}
Kyriakos Georgiou, Steve Kerrison, Zbigniew Chamski, and Kerstin Eder.
\newblock Energy transparency for deeply embedded programs.
\newblock {\em ACM Trans. Archit. Code Optim.}, 14(1), Mar 2017.

\bibitem[GR7]{GR712RCDevBoard}
{GR712RC Dual-Core LEON3-FT Development board}.
\newblock
  \url{https://www.gaisler.com/index.php/products/boards/gr712rc-board}.
\newblock Accessed: 2018-09-11.

\bibitem[GVGGS18]{guo2018things}
Xinfei Guo, Vaibhav Verma, Patricia Gonzalez-Guerrero, and Mircea~R Stan.
\newblock When “things” get older: exploring circuit aging in iot
  applications.
\newblock In {\em 2018 19th International Symposium on Quality Electronic
  Design (ISQED)}, pages 296--301. IEEE, 2018.

\bibitem[HWH95]{HWH95}
Christopher~A. Healy, David~B. Whalley, and Marion~G. Harmon.
\newblock Integrating the timing analysis of pipelining and instruction
  caching.
\newblock In {\em 16th {IEEE} Real-Time Systems Symposium, Palazzo dei
  Congressi, Via Matteotti, 1, Pisa, Italy, December 4-7, 1995, Proceedings},
  pages 288--297. {IEEE} Computer Society, 1995.

\bibitem[JML06]{Jayaseelan2006}
R.~Jayaseelan, T.~Mitra, and Xianfeng Li.
\newblock {Estimating the Worst-Case Energy Consumption of Embedded Software}.
\newblock In {\em Real-Time and Embedded Technology and Applications Symposium,
  2006. Proceedings of the 12th IEEE}, pages 81--90, April 2006.

\bibitem[KNNL05]{kutner2005applied}
Michael~H Kutner, Chris Nachtsheim, John Neter, and William Li.
\newblock {\em Applied linear statistical models}.
\newblock McGraw-Hill Irwin, 2005.

\bibitem[Leo]{Leon3FT}
{LEON3FT Fault-tolerant processor}.
\newblock \url{https://www.gaisler.com/index.php/products/processors/leon3ft}.
\newblock Accessed: 2018-09-11.

\bibitem[LH95]{lawson1995solving}
Charles~L Lawson and Richard~J Hanson.
\newblock {\em Solving least squares problems}.
\newblock SIAM, 1995.

\bibitem[MKE18]{Morse:2018}
Jeremy Morse, Steve Kerrison, and Kerstin Eder.
\newblock On the limitations of analyzing worst-case dynamic energy of
  processing.
\newblock {\em ACM Trans. Embed. Comput. Syst.}, 17(3):59:1--59:22, February
  2018.

\bibitem[NGC{\etalchar{+}}22]{nikov2022accurate}
Kris Nikov, Kyriakos Georgiou, Zbigniew Chamski, Kerstin Eder, and Jose
  Nunez-Yanez.
\newblock Accurate energy modelling on the cortex-m0 processor for profiling
  and static analysis.
\newblock In {\em 2022 29th IEEE International Conference on Electronics,
  Circuits and Systems (ICECS)}, pages 1--4. IEEE, 2022.

\bibitem[NMW{\etalchar{+}}22]{nikov2022robust}
Kris Nikov, Marcos Martinez, Simon Wegener, Jose Nunez-Yanez, Zbigniew Chamski,
  Kyriakos Georgiou, and Kerstin Eder.
\newblock Robust and accurate fine-grain power models for embedded systems with
  no on-chip pmu.
\newblock {\em IEEE Embedded Systems Letters}, 2022.

\bibitem[NNY20]{nikov2020intra}
Krastin Nikov and Jose Nunez-Yanez.
\newblock Intra and inter-core power modelling for single-isa heterogeneous
  processors.
\newblock {\em International Journal of Embedded Systems}, 12(3):324--340,
  2020.

\bibitem[NNYH15]{nikov2015evaluation}
Krastin Nikov, Jose~L Nunez-Yanez, and Matthew Horsnell.
\newblock Evaluation of hybrid run-time power models for the arm big. little
  architecture.
\newblock In {\em 2015 IEEE 13th International Conference on Embedded and
  Ubiquitous Computing}, pages 205--210. IEEE, 2015.

\bibitem[NYL13]{nunez2013enabling}
Jose Nunez-Yanez and Geza Lore.
\newblock Enabling accurate modeling of power and energy consumption in an
  arm-based system-on-chip.
\newblock {\em Microprocessors and Microsystems}, 37(3):319--332, 2013.

\bibitem[PHB13]{BEEBSb}
James Pallister, Simon~J. Hollis, and Jeremy Bennett.
\newblock {BEEBS:} open benchmarks for energy measurements on embedded
  platforms.
\newblock {\em CoRR}, abs/1308.5174, 2013.

\bibitem[PKME17]{pallister:2017}
James Pallister, Steve Kerrison, Jeremy Morse, and Kerstin Eder.
\newblock Data dependent energy modeling for worst case energy consumption
  analysis.
\newblock In {\em Proceedings of the 20th International Workshop on Software
  and Compilers for Embedded Systems}, SCOPES 2017, pages 51--59, New York, NY,
  USA, 2017. Association for Computing Machinery.

\bibitem[RAKK13]{rodrigues2013study}
Rance Rodrigues, Arunachalam Annamalai, Israel Koren, and Sandip Kundu.
\newblock A study on the use of performance counters to estimate power in
  microprocessors.
\newblock {\em IEEE Transactions on Circuits and Systems II: Express Briefs},
  60(12):882--886, 2013.

\bibitem[RBE{\etalchar{+}}22]{rouxel2022teamplay}
Benjamin Rouxel, Christopher~Mark Brown, Emad Ebeid, Kerstin Eder, Heiko Falk,
  Clemens Grelck, Jesper Holst, Shashank Jadhav, Yoann Marquer, Marcos Martinez
  De~Alejandro, et~al.
\newblock The teamplay project: analysing and optimising time, energy, and
  security for cyber-physical systems.
\newblock In {\em Proceedings the 2023 Design, Automation \& Test in Europe
  Conference \& Exhibition (DATE 2023)}. EDAA/IEEE, 2022.

\bibitem[RPBA{\etalchar{+}}14]{rethinagiri2014system}
Santhosh~Kumar Rethinagiri, Oscar Palomar, Rabie Ben~Atitallah, Smail Niar,
  Osman Unsal, and Adrian~Cristal Kestelman.
\newblock System-level power estimation tool for embedded processor based
  platforms.
\newblock In {\em Proceedings of the 6th Workshop on Rapid Simulation and
  Performance Evaluation: Methods and Tools}, pages 1--8, 2014.

\bibitem[SSEM21]{seewald2021coarse}
Adam Seewald, Ulrik~Pagh Schultz, Emad Ebeid, and Henrik~Skov Midtiby.
\newblock Coarse-grained computation-oriented energy modeling for heterogeneous
  parallel embedded systems.
\newblock {\em International Journal of Parallel Programming}, 49(2):136--157,
  2021.

\bibitem[STM]{STMFO_technical}
{STM32F030x4/x6/x8/xC and STM32F070x6/xB advanced ARM-based 32-bit MCUs -
  Reference Manual}.
\newblock Accessed: 2020-12-28.

\bibitem[TFW00]{TFW00}
Henrik Theiling, Christian Ferdinand, and Reinhard Wilhelm.
\newblock Fast and precise {WCET} prediction by separated cache and path
  analyses.
\newblock {\em Real-Time Systems}, 18(2/3):157--179, 2000.

\bibitem[Thu]{Thumbulator}
{The Thumbulator Git repository modified for collecting performance monitoring
  counters for the STM32F0-Discovery board.}
\newblock {branch: prefetch-model, Git-tag: teamplay-D4.5. Accessed:
  2020-11-09}.

\bibitem[WDD{\etalchar{+}}18]{wagemann2018whole}
Peter W{\"a}gemann, Christian Dietrich, Tobias Distler, Peter Ulbrich, and
  Wolfgang Schr{\"o}der-Preikschat.
\newblock Whole-system worst-case energy-consumption analysis for
  energy-constrained real-time systems.
\newblock {\em Leibniz International Proceedings in Informatics, LIPIcs 106
  (2018)}, 106:24, 2018.

\bibitem[WDH{\etalchar{+}}15]{wagemannworst}
P.~W{\"a}gemann, T.~Distler, T.~H{\"o}nig, H.~Janker, R.~Kapitza, and
  W.~Schr{\"o}der-Preikschat.
\newblock {Worst-Case Energy Consumption Analysis for Energy-Constrained
  Embedded Systems}.
\newblock In {\em 2015 27th Euromicro Conference on Real-Time Systems (ECRTS)},
  pages 105--114, July 2015.

\bibitem[WDH{\etalchar{+}}17]{Walker2017}
Matthew~J Walker, Stephan Diestelhorst, Andreas Hansson, Anup~K Das, Sheng
  Yang, Bashir~M Al-hashimi, and Geoff~V Merrett.
\newblock {Accurate and Stable Run-Time Power Modeling for Mobile and Embedded
  CPUs}.
\newblock {\em Ieee Transactions on Computer Aided Design of Integrated
  Circuits and Systems}, 36(1):1--14, 2017.

\bibitem[WEE{\etalchar{+}}08]{WEE+08}
Reinhard Wilhelm, Jakob Engblom, Andreas Ermedahl, Niklas Holsti, Stephan
  Thesing, David Whalley, Guillem Bernat, Christian Ferdinand, Reinhold
  Heckmann, Tulika Mitra, Frank Mueller, Isabelle Puaut, Peter Puschner, Jan
  Staschulat, and Per Stenstr{\"o}m.
\newblock The worst-case execution-time problem --- overview of methods and
  survey of tools.
\newblock {\em ACM Transactions on Embedded Computing Systems},
  7(3):36:1--36:53, May 2008.

\end{thebibliography}

\end{document}